\documentclass[aip,jcp,reprint,noshowkeys,superscriptaddress]{revtex4-2}
\usepackage{graphicx,dcolumn,bm,microtype,multirow,amscd,amsmath,amssymb,amsfonts,physics,wrapfig,bbold,siunitx,xspace,braket,txfonts,mathbbol}

\usepackage[version=4]{mhchem}

\usepackage[utf8]{inputenc}
\usepackage[T1]{fontenc}
\usepackage{feynmp-auto}
\usepackage{longtable}
\usepackage[dvipsnames]{xcolor}
\usepackage{mathtools}

\usepackage{hyperref}
\hypersetup{
    colorlinks,
    linkcolor={red!50!black},
    citecolor={red!70!black},
    urlcolor={red!80!black}
}

\usepackage{listings}
\definecolor{codegreen}{rgb}{0.58,0.4,0.2}
\definecolor{codegray}{rgb}{0.5,0.5,0.5}
\definecolor{codepurple}{rgb}{0.25,0.35,0.55}
\definecolor{codeblue}{rgb}{0.30,0.60,0.8}
\definecolor{backcolour}{rgb}{0.98,0.98,0.98}
\definecolor{mygray}{rgb}{0.5,0.5,0.5}

\definecolor{sqred}{rgb}{0.85,0.1,0.1}
\definecolor{sqgreen}{rgb}{0.25,0.65,0.15}
\definecolor{sqorange}{rgb}{0.90,0.50,0.15}
\definecolor{sqblue}{rgb}{0.10,0.3,0.60}

\lstdefinestyle{mystyle}{
    backgroundcolor=\color{backcolour},
    commentstyle=\color{codegreen},
    keywordstyle=\color{codeblue},
    numberstyle=\tiny\color{codegray},
    stringstyle=\color{codepurple},
    basicstyle=\ttfamily\footnotesize,
    breakatwhitespace=false,
    breaklines=true,
    captionpos=b,
    keepspaces=true,
    numbers=left,
    numbersep=5pt,
    numberstyle=\ttfamily\tiny\color{mygray},
    showspaces=false,
    showstringspaces=false,
    showtabs=false,
    tabsize=2
  }

  \newcolumntype{d}{D{.}{.}{-1}}

\usepackage{accents}

\newcommand{\ERI}[2]{\braket{#1|#2}}

\newcommand{\Om}{\Omega}

\newcommand{\titou}[1]{\textcolor{black}{#1}}

\newcommand{\bC}{\boldsymbol{C}}

\newcommand{\bX}{\boldsymbol{X}}
\newcommand{\bY}{\boldsymbol{Y}}

\newcommand{\bR}{\boldsymbol{R}}
\newcommand{\br}{\boldsymbol{r}}

\newcommand{\bM}{\boldsymbol{M}}
\newcommand{\bmm}{\boldsymbol{m}}
\newcommand{\bH}{\boldsymbol{H}}
\newcommand{\bh}{\boldsymbol{h}}
\newcommand{\bO}{\boldsymbol{0}}
\newcommand{\bI}{\boldsymbol{1}}

\newcommand{\bF}{\boldsymbol{F}}
\newcommand{\bff}{\boldsymbol{f}}

\newcommand{\bSig}{\boldsymbol{\Sigma}}
\newcommand{\bsig}{\boldsymbol{\sigma}}

\newcommand{\SupInf}{\textcolor{blue}{Supplementary Material}\xspace}
\newcommand{\mc}{\multicolumn}

\newcommand{\LCPQ}{Laboratoire de Chimie et Physique Quantiques (UMR 5626), Universit\'e de Toulouse, CNRS, Toulouse, France}
\newcommand{\UnHam}{Department of Chemistry, University of Hamburg, 22761 Hamburg, Germany; The Hamburg Centre for Ultrafast Imaging (CUI), Hamburg 22761, Germany}

\begin{document}	

\title{From Full Dynamic to Pure Static: A Family of $GW$-Based Approximations}

\author{Pierre-Fran\c{c}ois \surname{Loos}}
	\email{loos@irsamc.ups-tlse.fr}
	\affiliation{\LCPQ}

\author{Johannes \surname{T\"olle}}
        \email{johannes.toelle@uni-hamburg.de}
        \affiliation{\UnHam}
		
\begin{abstract}
We introduce a systematic hierarchy of one-body Green's function methods derived from the $GW$ approximation, constructed by progressively reducing the dynamical content of the self-energy.
Starting from the fully dynamical Dyson formulation, we generate a family of approximations that interpolates between the standard $GW$ approximation to purely static effective single-particle Hamiltonians.
This framework enables a controlled investigation of the role of dynamical effects and particle-hole coupling in the description of ionization potentials.
Within this unified formalism, the hole and particle branches can be selectively decoupled through downfolding strategies into reduced one-particle spaces.
By benchmarking the different members of this hierarchy on molecular ionization energies, we assess their accuracy, numerical robustness, and algorithmic complexity. 
We demonstrate that consistently derived partially static schemes can yield reliable quasiparticle energies while significantly simplifying the underlying eigenvalue problem.
We further introduce a novel static Hermitian self-energy obtained as the static limit of this hierarchy. 
Despite its conceptually distinct origin, it produces results remarkably close to those of qs$GW$, thereby providing an alternative static route toward partial self-consistency.
\end{abstract}

\maketitle

\section{Introduction}
\label{sec:intro}

The computation of charged excitations confronts electronic structure theory with a fundamental tension between physical fidelity and algorithmic tractability.
On the one hand, ionization potentials (IPs) and electron affinities (EAs) are governed by energy-dependent correlation effects stemming from the coupling between single-particle states and collective excitations.
On the other hand, practical calculations ultimately require the solution of well-conditioned eigenvalue problems, whose numerical properties are often at odds with the underlying many-body physics.
The history of Green's function methods for charged excitations can thus be read, to a large extent, as a sequence of compromises between these two requirements. \cite{Aryasetiawan_1998,Onida_2002,Reining_2017,Golze_2019,Marie_2024a,Blase_2018,Blase_2020,Krieger_2025}

Within this context, the $GW$ approximation \cite{Hedin_1965,Aryasetiawan_1998,Reining_2017,Golze_2019,Marie_2024a} has emerged as one of the most successful and widely used approaches for computing IPs and EAs, striking a favorable balance between accuracy and computational cost. \cite{Neuhauser_2014,Govoni_2015,Vlcek_2017,Wilhelm_2018,DelBen_2019,Forster_2020,Kaltak_2020,Forster_2021,Duchemin_2021,Wilhelm_2021,Forster_2022,Yu_2022,Tolle_2024}
Its popularity stems from a physically transparent description of dynamical screening, which provides an effective account of electron-electron interactions beyond mean-field theory, captures a large fraction of correlation effects relevant to charged excitations, \cite{Onida_2002,Martin_2016} and has well-defined formal connections with coupled-cluster theory. \cite{Lange_2018,Quintero_2022,Tolle_2023,Tolle_2025,Tolle_2025b,Kitsaras_2026,Coveney_2025a,Coveney_2025b,Coveney_2026,Tolle_2026} 

Despite its success, the $GW$ approximation admits a variety of practical implementations that differ significantly in how the self-energy is represented and evaluated.
In particular, most applications rely on a diagonal, perturbative treatment of the quasiparticle equation, commonly referred to as $G_0W_0$, \cite{Strinati_1980,Hybertsen_1985a,Godby_1988,Linden_1988,Northrup_1991,Blase_1994,Rohlfing_1995} in which off-diagonal couplings are neglected.
Closely related in spirit, the static Coulomb-hole plus screened-exchange (COHSEX) approximation \cite{Hedin_1965,Hybertsen_1986,Hedin_1999,Bruneval_2006} may be viewed as the zero-frequency limit of the $GW$ self-energy and provides a conceptually simple, albeit approximate, reference point within this hierarchy.

In addition, a broad spectrum of self-consistent extensions has been developed with the aim of reducing the dependence on the underlying mean-field starting point.
These include partially self-consistent schemes such as ev$GW$, \cite{Hybertsen_1986,Shishkin_2007a,Blase_2011a,Faber_2011,Marom_2012,Kaplan_2016,Wilhelm_2016,Rangel_2016} where only the quasiparticle energies \titou{entering the screened Coulomb interaction and the Green's function are updated}, and qs$GW$, \cite{Faleev_2004,vanSchilfgaarde_2006,Kotani_2007,Ke_2011,Gui_2018,Kaplan_2016,Forster_2021,Marie_2023,Forster_2025} which enforces self-consistency at the level of an effective static, Hermitian self-energy for both quasiparticle energies and orbitals.
At the opposite end of the spectrum lie fully self-consistent $GW$ approaches, in which the full frequency dependence (i.e., beyond the quasiparticle approximation) of both the Green's function and the self-energy is retained at each iteration. \cite{Caruso_2012,Caruso_2013a,Pokhilko_2021a,Pokhilko_2021b,Lei_2022,Harsha_2024,Wen_2024}
Together, these schemes highlight the flexibility of the $GW$ framework, but also underscore the lack of a unified computational approach.

A less frequently emphasized aspect of $GW$ implementations concerns the structure of the eigenvalue problem that must be solved to access physical observables.
In its most common diagonal form, $G_0W_0$ requires solving a nonlinear quasiparticle equation for each state of interest.
In contrast, in fully dynamical Dyson formulations, the self-energy can be represented through a supermatrix that couples the one-hole (1h) and one-particle (1p) states to the two-hole-one-particle (2h1p) and two-particle-one-hole (2p1h) configurations. \cite{Bintrim_2021,Monino_2022,Quintero_2022,Monino_2023,Tolle_2023,Scott_2023,Marie_2023}
In this coupled representation, the physically relevant quasiparticle energies, namely the principal IPs and EAs, appear as interior eigenvalues of a large matrix.
Accessing these states reliably, therefore, requires dedicated algorithms, such as state-following variants of the Davidson method, which allow one to track specific eigenstates.
Although less robust, these techniques also make it possible, in principle, to access inner-valence IPs states.

Between these two extremes lies a broad landscape of intermediate approximations.
In particular, it is not a priori necessary to treat all components of the self-energy on the same dynamical footing.
The hole and particle branches of the $GW$ self-energy encode distinct physical processes and may be selectively downfolded or rendered static without entirely abandoning dynamical correlation effects.
Such asymmetric treatments naturally give rise to hybrid schemes that interpolate between fully dynamical Dyson formulations and purely static effective Hamiltonians.

In parallel, a number of non-Dyson approaches have been explored, most notably within the algebraic-diagrammatic construction (ADC) framework, \cite{Schirmer_1982,Schirmer_1983,Schirmer_1984,Schirmer_2018,Trofimov_2005,Dreuw_2023} where the explicit coupling between hole and particle sectors is removed by construction. \cite{Schirmer_1998,Trofimov_2002,Schirmer_1991,Mertins_1996a,Mertins_1996b,Dempwolff_2019,Leitner_2024}
A notable advantage of the non-Dyson framework is that the IP and EA sectors can be explicitly decoupled. \cite{Schirmer_2018}
In this block-separated formulation, the principal IPs and EAs become extremal eigenvalues of their respective blocks.
This transformation eliminates the need to target interior eigenvalues and enables the use of standard Davidson algorithms.

Recent studies have reported that non-Dyson schemes within the $GW$ approximation may exhibit deceptive or inconsistent performance compared to their Dyson counterparts. \cite{Bintrim_2021,Tolle_2023}
These observations raise fundamental questions regarding the role of dynamical effects and sector coupling in Green's function methods.
In this work, we revisit these issues from a unified perspective.
Starting from the supermatrix formulation of the $GW$ approximation, we introduce a systematic hierarchy of schemes obtained by progressively reducing the dynamical content of the self-energy.
By selectively downfolding the 2h1p and 2p1h configuration spaces and by evaluating specific branches of the self-energy in the static limit, we construct a continuous family of approximations ranging from the fully dynamical $GW$ theory to purely static effective single-particle Hamiltonians.

This framework allows us to disentangle, in a systematic manner, the effects of dynamical correlations and sector coupling.
It also clarifies the formal connections between standard $G_0W_0$ practice, half-dynamical hybrid schemes, and approaches mimicking a non-Dyson construction, reminiscent of ADC-like constructions.
Through numerical benchmarks on molecular IPs, we assess the accuracy and limitations of the various members of this hierarchy and demonstrate that these schemes can perform significantly better than previously anticipated when derived consistently.

\section{Theory}
\label{sec:theory}
In this work, the indices $p,q,r,s, \dots$ are used for arbitrary spin orbitals, $i,j,k,l$ label the occupied orbitals, $a,b,c,d$ denote the virtual orbitals, and $\mu, \nu, \dots $ (collective) combined particle-hole indices.
The 1h space has the size of the number of occupied orbitals and will be denoted $N^\text{1h}$, while the 1p space has the size of the number of unoccupied orbitals and will be denoted $N^\text{1p}$.
We also define the size as the 2h1p space and the 2p1h space has $N^\text{2h1p} = (N^\text{1h})^2 N^\text{1p}$ and $N^\text{2p1h} = N^\text{1h} (N^\text{1p})^2$, respectively.
In the following, the theory section is structured according to the target single-particle space to which the decoupling is applied: the full single-particle space (1h+1p, Sec.~\ref{sec:dyson}), a reduced single-particle space (1h or 1p only, Sec.~\ref{sec:non-dyson}), and diagonal elements only (Sec.~\ref{sec:diag}).

\subsection{Full single-particle space (1h+1p)}
\label{sec:dyson}

The construction of the different members of the present hierarchy starts from a reformulation of the $GW$ approximation (used here synonymously with $G_0W_0$) as a linear eigenvalue problem 
\begin{equation}\label{eq:e_super_matrix}
	\qty( \bH - \omega \bI ) \cdot \bR = \bO
\end{equation}
known as the supermatrix representation. \cite{Bintrim_2021,Monino_2022,Quintero_2022,Monino_2023,Tolle_2023,Scott_2023,Marie_2023}
This formulation makes explicit the coupling between single-particle states and higher-order particle-hole configurations.
In this representation, the effective Hamiltonian reads
\begin{equation} \label{eq:super_matrix}
	\bH = 
	\begin{pmatrix}
		\bF		&	\qty(\bM^{\text{2h1p}})^{\dag}	&	\qty(\bM^{\text{2p1h}})^{\dag}
		\\
		\bM^{\text{2h1p}}	&	\bC^{\text{2h1p}}			&	\bO
		\\
		\bM^{\text{2p1h}}	&	\bO				&	\bC^{\text{2p1h}}	
	\end{pmatrix}
\end{equation}
where the Fock matrix $\bF$ is conveniently decomposed into its 1h and 1p subspaces 
\begin{equation} 
	\bF = 
	\begin{pmatrix}
		\bff^\text{1h}		&	\bO
		\\
		\bO	&	\bff^\text{1p}	
	\end{pmatrix}
\end{equation}
(In other words, Brillouin's theorem is fulfilled.)
Throughout this work, we assume a canonical Hartree--Fock (HF) orbital basis, which renders the Fock matrix diagonal.
The corresponding eigenvector can be decomposed in an analogous manner according to its components in the different subspaces
\begin{align} 
	\bR & = 
	\begin{pmatrix}
		\br		
		\\
		\br^{\text{2h1p}}
		\\
		\br^{\text{2p1h}}
	\end{pmatrix}
	&
	\br = 
	\begin{pmatrix}
		\br^{\text{1h}}
		\\
		\br^{\text{1p}}
	\end{pmatrix}	
\end{align}
This linear eigenvalue problem can equivalently be recast as a nonlinear quasiparticle equation
\begin{equation} \label{eq:quasiparticle_equation}
	\qty[ \bF + \bSig(\omega) - \omega \bI ] \cdot \br = \bO 
\end{equation}
where the correlation part of the self-energy is given by
\begin{equation} \label{eq:self_energy}
\begin{split}
	\bSig(\omega) 
	& = \bSig^\text{2h1p}(\omega) + \bSig^\text{2p1h}(\omega)
	\\
	& = \qty( \bM^{\text{2h1p}} )^{\dag} \cdot \qty(\omega \bI - \bC^{\text{2h1p}} )^{-1} \cdot \bM^{\text{2h1p}}
	\\
	& + \qty( \bM^{\text{2p1h}} )^{\dag} \cdot \qty(\omega \bI - \bC^{\text{2p1h}} )^{-1} \cdot \bM^{\text{2p1h}}
\end{split}
\end{equation}
The spectral weight of a given solution is given by $Z = \norm{\br}^2$.
Solving the non-linear equations \eqref{eq:quasiparticle_equation} is completely analogous to solving the linear eigenproblem \eqref{eq:super_matrix}.
Both formulations yield the same set of eigenvalues, corresponding to quasiparticle energies as well as satellite states.
When the quasiparticle solutions can be well identified from the satellite states, one gets $N^\text{1h} + N^\text{1p}$ quasiparticle solutions and $N^\text{2h1p} + N^\text{2p1h}$ satellite solutions.

As readily seen in Eq.~\eqref{eq:self_energy}, the self-energy thus naturally separates into a hole (or 2h1p) branch and a particle (or 2p1h) branch.
This formulation corresponds to a full downfolding of the 2h1p and 2p1h configuration spaces onto the combined 1h+1p space. 
The resulting self-energy $\bSig(\omega)$ is therefore \textit{fully dynamical}, as both branches retain an explicit frequency dependence.
Moreover, since the 1h and 1p sectors are coupled through $\bSig(\omega)$, this formulation is referred to as 1h+1p, or Dyson scheme.

The matrix elements of the fully dynamical self-energy read
\begin{equation}
\begin{split}
	\Sigma_{pq}(\omega) 
	& = \Sigma_{pq}^\text{2h1p}(\omega) 
	+ \Sigma_{pq}^\text{2p1h}(\omega) 
	\\
	& = \sum_{k\nu} \frac{M_{k\nu,p}M_{k\nu,q}}{\omega - \epsilon_{k} + \Om_{\nu}}
	+ \sum_{c\nu} \frac{M_{c\nu,p}M_{c\nu,q}}{\omega - \epsilon_{c} - \Om_{\nu}}
\end{split}
\end{equation}
where the coupling blocks are
\begin{align}
	M^{\text{2h1p}}_{k\nu,q} & = M_{k\nu,q}
	&
	M^{\text{2p1h}}_{c\nu,q} & = M_{c\nu,q}
\end{align}
with
\begin{equation}
	M_{p\nu,q} = \sum_{kc} \ERI{pc}{qk} (\bX+\bY)_{kc,\nu} 
\end{equation}
and the $\bC$ blocks are diagonal with elements
\begin{align}
	C^{\text{2h1p}}_{k\nu,k\nu} & = \epsilon_i - \Omega_{\nu}
	&
	C^{\text{2p1h}}_{c\nu,c\nu} & = \epsilon_a + \Omega_{\nu}
\end{align}
Here, $\epsilon_p$ are the Hartree--Fock orbital energies, $\ERI{pq}{rs}$ is a two-electron integral in Dirac notation, $\Omega_{\nu}$ is a (positive) random-phase approximation (RPA) excitation energy, and $(\bX+\bY)_\nu$ its corresponding eigenvector.

While the fully dynamic formulation provides a formally exact reformulation of the $GW$ problem within the chosen configuration space, it is not the only possible representation. 
In practice, alternative downfolding strategies may be advantageous, either to reduce computational cost or to facilitate the construction of static, Hermitian effective Hamiltonians. 
In particular, one may choose to downfold only part of the configuration space, leading to hybrid representations in which some branches of the self-energy remain frequency dependent while others are made static. 
Such constructions naturally interpolate between fully dynamic Dyson schemes and static quasiparticle approximations.

One such possibility consists in downfolding only the 2p1h sector onto the 1h+1p space, while retaining the explicit coupling to the 2h1p configurations. 
This choice leads to the following reduced (frequency-dependent) supermatrix representation:
\begin{equation} 
	\bH^\text{2h1p}(\omega) = 
	\begin{pmatrix}
		\bF + \bSig^\text{2p1h}(\omega)		&	\qty(\bM^{\text{2h1p}})^{\dag}	
		\\
		\bM^{\text{2h1p}}					&	\bC^{\text{2h1p}}	
	\end{pmatrix}
\end{equation}
Importantly, as long as the full frequency dependence of the 2p1h self-energy is retained, this representation is formally equivalent to the fully dynamic formulation discussed above and yields exactly the same eigenvalues.
To obtain a Hermitian matrix with a static 1h+1p block, one may instead evaluate the 2p1h self-energy elements  $\bSig_{pq}^\text{2p1h}(\omega)$ at the symmetric energy point $\omega = (\epsilon_p + \epsilon_q)/2$.
This yields
\begin{equation} \label{eq:SigInf_2p1h}
	\Bar{\Sigma}_{pq}^\text{2p1h} 
	= \sum_{c\nu} \frac{M_{c\nu,p}M_{c\nu,q}}{(\epsilon_p + \epsilon_q)/2 - \epsilon_{c} - \Om_{\nu}}
\end{equation}
and the corresponding quasiparticle equation becomes
\begin{equation}
	\qty[ \bF + \bSig^\text{2h1p}(\omega) + \Bar{\bSig}^\text{2p1h} - \omega \bI ] \cdot \br = \bO 
\end{equation}
In this construction, the hole branch remains frequency dependent, while the particle branch is made static.
As a result, the spectrum is no longer identical to that of the fully dynamic self-energy, and both quasiparticle and satellite energies are modified.
For brevity, we refer to this approximation as the \textit{half-and-half} (h\&h) self-energy.
In Appendix \ref{app:ADC_choice}, we discuss an alternative choice for a symmetric static branch based on the non-Dyson ADC scheme. \cite{Schirmer_2018} Numerical experiments show that the quasiparticle energies are largely insensitive to the specific choice of symmetrization.

This construction highlights that the frequency dependence of the self-energy is not an all-or-nothing feature, but can instead be selectively retained or eliminated in a systematic manner. 
When the full frequency dependence of the downfolded branch is preserved, the resulting formulation remains exactly equivalent to the fully dynamic Dyson scheme.
Approximations arise only once the self-energy is evaluated at a fixed frequency, which effectively replaces a dynamical correlation effect by an effective static one.

The resulting h\&h self-energy thus provides a well-defined intermediate approximation, in which dynamical correlation effects are treated asymmetrically between hole and particle sectors. 
From a physical perspective, this approximation amounts to retaining satellite structure on one side of the spectrum while collapsing the opposite branch into an effective static potential. 
Such a construction is particularly appealing in contexts where either the hole or particle satellites are known to play a dominant role. \cite{Lundqvist_1969,Langreth_1970,Gunnarsson_1994,Hedin_1980,Hedin_1999,Loos_2024}
\titou{Importantly, within the matrix formulation, the principal IP (typically the quantity of greatest interest) becomes the lowest-energy solution of the associated linear system, thereby enabling the use of efficient iterative eigensolvers such as the Davidson algorithm, as routinely done in non-Dyson ADC approaches (see below). \cite{Schirmer_1998,Trofimov_2005,Schirmer_2018}}

A fully analogous construction can be carried out for the particle sector by instead downfolding the 2h1p branch
\begin{equation} 
	\Bar{\Sigma}_{pq}^\text{2h1p}
	= \sum_{k\nu} \frac{M_{k\nu,p}M_{k\nu,q}}{(\epsilon_p + \epsilon_q)/2 - \epsilon_{k} + \Om_{\nu}}
\end{equation}

Finally, downfolding both branches yields a purely static self-energy, 
\begin{equation}
	\Bar{\bSig} = \Bar{\bSig}^\text{2h1p} + \Bar{\bSig}^\text{2p1h}
\end{equation}
which can be viewed as an effective single-particle Hamiltonian incorporating correlation effects beyond Hartree--Fock:
\begin{equation}
	\qty[ \bF + \Bar{\bSig}^\text{2h1p} + \Bar{\bSig}^\text{2p1h} - \omega \bI ] \cdot \br = \bO 
\end{equation} 
While this limit sacrifices the explicit description of satellite features, it provides a simple and Hermitian framework that may be useful for exploratory studies or as a starting point for partially self-consistent schemes (see Sec.~\ref{sec:res}). 
Within the present hierarchy, these various choices naturally define a family of approximations with increasing dynamical content, all derived from the same underlying formalism.

\subsection{Reduced single-particle space, 1h or 1p only}
\label{sec:non-dyson}

We now turn to the construction of a variant restricted to the reduced single-particle space (1h or 1p only).
Loosely speaking, this approach can be categorized as a non-Dyson approach.
The central idea is to decouple the 1h and 1p sectors by eliminating their explicit dynamical coupling, while still retaining a frequency-dependent description of the dominant correlation effects associated with shake-up processes. 
In contrast to a Dyson scheme, where both sectors are mixed through the self-energy, the non-Dyson construction aims at a block-diagonal structure in which hole and particle excitations can be treated independently. \cite{Schirmer_1998,Trofimov_2005,Schirmer_2018}

We start by restricting the internal space to the 1h sector of the Fock matrix. 
In this case, the supermatrix representation of the problem reads
\begin{equation} \label{eq:dyn_1h_sector}
	\bh = 
	\begin{pmatrix}
		\bff^\text{1h}		&	\qty(\bmm^{\text{2h1p}})^{\dag}	&	\qty(\bmm^{\text{2p1h}})^{\dag}
		\\
		\bmm^{\text{2h1p}}	&	\bC^{\text{2h1p}}				&	\bO
		\\
		\bmm^{\text{2p1h}}	&	\bO								&	\bC^{\text{2p1h}}	
	\end{pmatrix}
\end{equation}
where $\bmm^{\text{2h1p}}$ is the occupied block of $\bM^{\text{2h1p}}$, that is, $m_{k\nu,i}^{\text{2h1p}} = M_{k\nu,i}^{\text{2h1p}}$.

A full downfolding of both the 2h1p and 2p1h configuration spaces onto the 1h subspace leads to the following fully-dynamical self-energy:
\begin{equation} 
\begin{split} 
	\bsig(\omega) 
	& = \bsig^\text{2h1p}(\omega) + \bsig^\text{2p1h}(\omega)
	\\
	& = \qty(\bmm^{\text{2h1p}})^{\dag}  \cdot \qty(\omega \bI - \bC^{\text{2h1p}} )^{-1} \cdot \bmm^{\text{2h1p}}
	\\
	& + \qty(\bmm^{\text{2p1h}})^{\dag}  \cdot \qty(\omega \bI - \bC^{\text{2p1h}} )^{-1} \cdot \bmm^{\text{2p1h}}
\end{split}
\end{equation}
with matrix elements
\begin{equation}
\begin{split}
	\sigma_{ij}(\omega) 
	& = \sigma_{ij}^\text{2h1p}(\omega) + \sigma_{ij}^\text{2p1h}(\omega)
	\\
	& = \sum_{k\nu} \frac{M_{k\nu,i}M_{k\nu,j}}{\omega - \epsilon_{k} + \Om_{\nu}}
	+ \sum_{c\nu} \frac{M_{c\nu,i}M_{c\nu,j}}{\omega - \epsilon_{c} - \Om_{\nu}}
\end{split}
\end{equation}
The corresponding quasiparticle equation reads
\begin{equation} 
	\qty[ \bff^\text{1h} + \bsig(\omega) - \omega \bI ] \cdot \br^\text{1h} = \bO 
\end{equation}
Despite the restriction of the internal space to 1h configurations, this formulation does not yet constitute a genuine non-Dyson scheme, as the hole sector remains dynamically coupled to the particle sector through the frequency-dependent contribution $\bsig^\text{2p1h}(\omega)$.

To achieve a true decoupling of the 1h and 1p sectors, we therefore only downfold the 2p1h configurations onto the 1h block, yielding the reduced  (frequency-dependent) supermatrix
\begin{equation}\label{eq:1h_hh_scheme}
	\bh^\text{2h1p}(\omega) = 
	\begin{pmatrix}
		\bff^\text{1h} + \bsig^\text{2p1h}(\omega)		&	\qty(\bmm^{\text{2h1p}})^{\dag}	
		\\
		\bmm^{\text{2h1p}}	&	\bC^{\text{2h1p}}	
	\end{pmatrix}
\end{equation}
In practice, in order to keep the effective 1h block Hermitian and frequency independent, the 2p1h self-energy is evaluated at a symmetric energy point, $\omega = (\epsilon_i + \epsilon_j)/2$. 
This leads to the following expression for the non-Dyson self-energy in the hole sector:
\begin{equation}
	\Bar{\sigma}_{ij}^\text{2p1h}
	= \sum_{c\nu} \frac{M_{c\nu,i}M_{c\nu,j}}{(\epsilon_i + \epsilon_j)/2 - \epsilon_{c} - \Om_{\nu}}
\end{equation}
and the corresponding quasiparticle equation becomes
\begin{equation}
	\qty[ \bff^\text{1h} + \bsig^\text{2h1p}(\omega) + \Bar{\bsig}^\text{2p1h} - \omega \bI ] \cdot \br^\text{1h} = \bO 
\end{equation}
In this formulation, the hole (2h1p) branch retains its explicit frequency dependence, while the particle (2p1h) contribution is rendered static.
An entirely analogous construction can be carried out for the particle sector, leading to
\begin{equation}
	\Bar{\sigma}_{ij}^\text{2h1p}
	= \sum_{k\nu} \frac{M_{k\nu,i}M_{k\nu,j}}{(\epsilon_i + \epsilon_j)/2 - \epsilon_{k} + \Om_{\nu}}
\end{equation}

Finally, by evaluating both branches in the static limit, one obtains a purely static matrix in the 1h (or 1p) space
\begin{equation} \label{eq:1h+1p_static}
	\Bar{\bsig} = \Bar{\bsig}^\text{2h1p} + \Bar{\bsig}^\text{2p1h}
\end{equation}
which corresponds to a complete removal of dynamical effects from the self-energy:
\begin{equation}
	\qty[ \bff^\text{1h} + \Bar{\bsig}^\text{2h1p} + \Bar{\bsig}^\text{2p1h} - \omega \bI ] \cdot \br^\text{1h} = \bO 
\end{equation} 

\subsection{Diagonal approximation}
\label{sec:diag}
The constructions introduced in the two preceding subsections can be further simplified by restricting the analysis to a single diagonal element of the 1h block of the Fock matrix. 
This corresponds to neglecting off-diagonal couplings within the 1h space and treating each orbital independently.
Such a reduction leads naturally to the diagonal approximation, which is commonly employed in practical $G_0W_0$ calculations. \cite{Marie_2024a}

Within this approximation, the problem associated with a given occupied orbital $i$ can be formulated in terms of the following reduced supermatrix:
\begin{equation} \label{eq:diagonal_approximation}
	\bh_i = 
	\begin{pmatrix}
		\epsilon_{i}			&	\qty(\bmm_i^{\text{2h1p}})^{\dag}	&	\qty(\bmm_i^{\text{2p1h}})^{\dag}
		\\
		\bmm_i^{\text{2h1p}}	&	\bC^{\text{2h1p}}					&	\bO
		\\
		\bmm_i^{\text{2p1h}}	&	\bO									&	\bC^{\text{2p1h}}	
	\end{pmatrix}
\end{equation}
where $\bmm_i^{\text{2h1p}}$ and $\bmm_i^{\text{2p1h}}$ denote the coupling vectors between the reference hole state $i$ and the corresponding 2h1p and 2p1h configuration spaces.
Downfolding these configurations onto the single 1h state yields a scalar, frequency-dependent self-energy,
\begin{equation} 
\begin{split} 
	\sigma_{ii}(\omega) 
	& = \sigma_{ii}^\text{2h1p}(\omega) + \sigma_{ii}^\text{2p1h}(\omega)
	\\
	& = \qty(\bmm_i^{\text{2h1p}})^{\dag}  \cdot \qty(\omega \bI - \bC^{\text{2h1p}} )^{-1} \cdot \bmm_i^{\text{2h1p}}
	\\
	& + \qty(\bmm_i^{\text{2p1h}})^{\dag}  \cdot \qty(\omega \bI - \bC^{\text{2p1h}} )^{-1} \cdot \bmm_i^{\text{2p1h}}
\end{split}
\end{equation}
which is precisely the form of the diagonal $GW$ self-energy used in conventional calculations.

In line with the hierarchical constructions introduced above, several levels of approximation can be defined depending on how the hole and particle branches of the self-energy are treated.
In particular, one may retain the full dynamical structure of both branches, render one of them static, or evaluate both in the static limit.
Within the diagonal approximation, this leads to a corresponding hierarchy of diagonal self-energies, constructed from the following static contributions:\begin{subequations}
\begin{align}
	\Bar{\sigma}_{ii}^\text{2h1p} & = \sum_{k\nu} \frac{M_{k\nu,i}^2}{\epsilon_i - \epsilon_{k} + \Om_{\nu}}
	\\
	\Bar{\sigma}_{ii}^\text{2p1h} & = \sum_{c\nu} \frac{M_{c\nu,i}^2}{\epsilon_i - \epsilon_{c} - \Om_{\nu}} \label{eq:1h_approx}
	\\
	\Bar{\sigma}_{ii} & = \Bar{\sigma}_{ii}^\text{2h1p} + \Bar{\sigma}_{ii}^\text{2p1h}
\end{align}
\end{subequations}
The first two expressions define diagonal variants in which only one branch of the self-energy is treated dynamically, while the other is replaced by its static counterpart.
The last expression corresponds to the pure static diagonal approximation, in which all dynamical effects are removed from the self-energy.

These diagonal limits can thus be viewed as the simplest members of the broader hierarchy introduced in this work, and they offer a transparent connection to standard $G_0W_0$ practice while highlighting possible systematic extensions beyond it.

\section{Results and Discussion}
\label{sec:res}

To evaluate the performance of the different schemes, we compute the inner- and outer-valence IPs for the molecular benchmark set reported in Ref.~\onlinecite{Marie_2024b}.
This dataset gathers 58 valence IPs for small molecular systems, evaluated in the aug-cc-pVTZ basis. 
The resulting values are compared with (i) fully dynamical $GW$ results obtained within the complete 1h+1p reference space as defined in Eq.~\eqref{eq:e_super_matrix} (Table \ref{tab:IPwrtGW}), and (ii) theoretical best estimates (TBEs) extracted from Ref.~\onlinecite{Marie_2024b}, computed at the full configuration interaction (FCI) level (Table \ref{tab:IPwrtTBE}).
For each case, the mean-absolute error (MAE), mean-signed error (MSE), and the absolute maximum error (Max) are reported.
The raw data can be found in the \SupInf.
\titou{Unless otherwise stated, all calculations are performed within the one-shot $G_0W_0$ framework, i.e., without self-consistency.}

From Table \ref{tab:IPwrtGW}, we see that the difference between the full dynamical 1h+1p treatment and the diagonal approximation [see Eq.~\eqref{eq:diagonal_approximation}] is found to be very small, with a MAE of \SI{0.015}{\eV}, a MSE of \SI{-0.007}{\eV}, and a maximum absolute error of \SI{0.118}{\eV}. 
These results confirm that the widely used diagonal approximation is well justified in this context.
Note, however, that the current comparison is based on canonical Hartree--Fock orbitals, although similar findings have also been reported for other mean-field orbitals. \cite{Kaplan_2015}
Restricting the reference space to 1h only [see Eq.~\eqref{eq:dyn_1h_sector}] proves to be an even more accurate approximation, yielding a systematic (though very small) underestimation of the IPs, with MAE and MSE values of \SI{-0.012}{\eV} relative to the 1h+1p treatment.

This naturally raises the question of whether the frequency dependence of the self-energy can be further reduced without compromising accuracy. 
In this respect, the (non-Dyson) 1h-h\&h scheme [see Eq.~\eqref{eq:1h_hh_scheme}] appears particularly promising, yielding a MAE of \SI{0.021}{\eV}, a MSE of \SI{-0.021}{\eV}, and a maximum absolute error of only \SI{0.075}{\eV}.

By contrast, the alternative scheme exhibits somewhat poorer overall statistics and, more importantly, pronounced outliers, with maximum errors exceeding \SI{1}{\eV} in certain cases. 
In the present work, however, we demonstrate that these large deviations can be substantially mitigated through the use of a regularization procedure. 
The occurrence of such outliers can be traced back to numerical instabilities in the evaluation of the self-energy rather than to an intrinsic failure of the underlying approximation.

We have thus reported the same statistical indicators as above, now obtained using a SRG-based regularization with a flow parameter of $s = 500$, instead of the essentially unregularized choice $s = \num{d6}$ employed previously. \cite{Marie_2023}
Within the SRG framework, potentially divergent terms of the form $x^{-1}$ arising near $x = 0$ are replaced by the regularized expression $x^{-1}(1 - e^{-s x^2})$, which remains well behaved in this limit. \cite{Evangelista_2014b,Hergert_2016,ChenyangLi_2019a,Krieger_2025}
Further technical details on the regularization procedure can be found in Refs.~\onlinecite{Marie_2023,Marie_2025b}.

\begin{table*}
\caption{Mean-absolute, mean-signed, and maximum absolute errors (MAE/MSE/Max), in \si{\eV}, computed for various schemes with respect to the $GW$ fully-dynamical results considering the entire 1h+1p reference space. All calculations are performed with the aug-cc-pVTZ basis.}
\label{tab:IPwrtGW}
\begin{ruledtabular}
\begin{tabular}{c|cccccc}
Reference	&	\mc{3}{c}{Self-energy ($s = \num{d6}$)}						&	\mc{3}{c}{Self-energy ($s = 500$)}												\\
				\cline{2-4} \cline{5-7}
space		&	full dynamic	&	h\&h	&	pure static		&	full dynamic	&	h\&h	&	pure static		\\	
\hline
1h+1p		&	0.000/0.000/0.000	&	0.095/0.089/1.695	&	0.179/-0.001/1.827	&	0.000/0.000/0.000	&	0.014/0.001/0.076	&	0.104/-0.089/0.875	\\	
1h			&	0.012/-0.012/0.074	&	0.021/-0.021/0.075	&	0.122/-0.118/1.405	&						&	0.021/-0.021/0.075	&	0.112/-0.109/0.878	\\
diagonal	&	0.015/-0.007/0.118	&	0.040/-0.017/0.554	&	0.127/-0.104/1.405	&	0.015/-0.007/0.118	&	0.021/-0.017/0.131	&	0.103/-0.099/0.878	\\
\end{tabular}
\end{ruledtabular}
\end{table*}

\begin{table*}
\caption{Mean-absolute, mean-signed, and maximum absolute errors (MAE/MSE/Max), in \si{\eV}, computed for various schemes with respect to the theoretical best estimates. All calculations are performed with the aug-cc-pVTZ basis.}
\label{tab:IPwrtTBE}
\begin{ruledtabular}
\begin{tabular}{c|cccccc}
Reference	&	\mc{3}{c}{Self-energy ($s = \num{d6}$)}						&	\mc{3}{c}{Self-energy ($s = 500$)}									\\
				\cline{2-4} \cline{5-7}
space		&	full dynamic	&	h\&h	&	pure static		&	full dynamic	&	h\&h	&	pure static		\\	
\hline
1h+1p		&	0.453/0.406/2.063	&	0.526/0.495/2.069	&	0.505/0.405/1.741	&	0.453/0.406/2.063	&	0.452/0.407/2.046	&	0.415/0.318/1.756	\\	
1h			&	0.451/0.394/2.057	&	0.445/0.385/2.036	&	0.422/0.288/1.744	&						&	0.445/0.385/2.036	&	0.411/0.297/1.751	\\
diagonal	&	0.453/0.399/2.052	&	0.464/0.390/2.032	&	0.447/0.303/1.744	&	0.453/0.399/2.052	&	0.447/0.390/2.032	&	0.416/0.308/1.749	\\
\end{tabular}
\end{ruledtabular}
\end{table*}

A clear distinction emerges from Table \ref{tab:IPwrtGW} between intrinsic approximation errors and numerical instabilities associated with the evaluation of the self-energy. 
This separation becomes possible through the use of the SRG-based regularization, which selectively affects schemes involving problematic energy denominators while leaving numerically stable formulations unchanged.

The fully dynamical schemes do not require regularization; their results are therefore unchanged.
A markedly different behavior is observed for the partially dynamical h\&h scheme. 
Without regularization ($s=\num{d6}$), this approximation exhibits apparently poor statistics in the full 1h+1p reference space, with a MAE of \SI{0.095}{\eV} and very large outliers reaching \SI{1.695}{\eV}. 
However, once the SRG regularization is introduced ($s=500$), the errors collapse dramatically: the MAE drops to \SI{0.014}{\eV} and the maximum deviation to only \SI{0.076}{\eV}. 
A similar stabilization is observed in the diagonal and 1h subspaces, where the h\&h scheme consistently yields errors on the order of a few hundredths of an eV. 
When these pathologies are removed, the h\&h approximation proves to be nearly as accurate as the fully dynamical treatment, despite its reduced frequency dependence.

The situation is qualitatively different for the purely static approximation. 
Regularization again suppresses extreme outliers (e.g., reducing the maximum error in the 1h+1p space from \SI{1.827}{\eV} to \SI{0.875}{\eV}), but the MAEs remain substantial, on the order of \SI{0.10}{\eV}. 
This indicates that, in this case, the dominant source of error is not numerical but physical: the complete removal of dynamical effects in both branches of the self-energy leads to a genuine loss of accuracy that cannot be recovered through regularization alone.

Finally, the diagonal approximation follows the same overall pattern. 
For the h\&h variant, SRG regularization further improves the results and eliminates the larger deviations observed in the unregularized case, while the purely static diagonal scheme remains significantly less accurate. 
The consistent behavior across all reference spaces reinforces the conclusion that SRG acts as a targeted numerical remedy rather than as an empirical correction to the physics. \cite{Evangelista_2014b,Hergert_2016}

Taken together, these results show that the SRG regularization serves primarily to expose the true performance of the underlying approximations by removing spurious divergences. 
Once these numerical artifacts are controlled, the h\&h scheme emerges as a remarkably accurate representation of the fully dynamical $GW$ self-energy for valence IPs, whereas the purely static scheme remains limited by more fundamental physical approximations.
However, owing to a fortuitous cancellation of errors, the static schemes produce IPs that are slightly closer to the reference TBE values, with improvements of approximately \SI{0.04}{\eV} in MAE and \SI{0.10}{\eV} in MSE (see Table \ref{tab:IPwrtTBE}).

Within this same regularized framework, we next assess the performance of the present static \textit{self-consistent} scheme relative to qs$GW$. 
For the same set of IPs and basis set, qs$GW$ yields a MAE of \SI{0.336}{\eV} and a MSE of \SI{0.290}{\eV} (see \SupInf), which provides a useful reference point for assessing the performance of related self-consistent static schemes. 
In this spirit, we have implemented a self-consistent scheme based on the purely static self-energy of Eq.~\eqref{eq:1h+1p_static} and compared its performance to qs$GW$. 
For both schemes, we employ the same SRG regularization procedure with a flow parameter $s = 500$.
This approach results in a MAE of \SI{0.330}{\eV} and a MSE of \SI{0.278}{\eV}, representing a marginal improvement over qs$GW$, yet a substantial one relative to its one-shot counterpart.

In qs$GW$, the static effective self-energy is defined as \cite{Kotani_2007,Ismail-Beigi_2017} 
\begin{equation} \label{eq:sigmatilde}
	\Sigma_{pq}^\text{qs$GW$} = \frac{\Sigma_{pq}(\epsilon_p) + \Sigma_{qp}(\epsilon_q)}{2}
\end{equation}
This construction corresponds to the usual qs$GW$ static self-energy.
\footnote{The definition of Eq.~\eqref{eq:sigmatilde} corresponds to the so-called ``mode A'' introduced by Kotani, van Schilfgaarde, and Faleev. \cite{Kotani_2007} 
In the same work, the authors proposed an alternative formulation, denoted ``mode B'', which is frequently employed in qs$GW$ implementations relying on analytical continuation of the self-energy. \cite{Lei_2022,Harsha_2024,Forster_2025} 
In mode B, the off-diagonal elements of the self-energy are evaluated at the Fermi energy, a choice that is numerically more stable than evaluating them at energies far from this reference. 
This modification generally improves convergence while producing results that are qualitatively and quantitatively comparable to those of mode A. Nevertheless, since the expression in Eq.~\eqref{eq:sigmatilde} remains the more widely used definition, the present work is restricted to mode A.}
In this approach, the static operator is obtained by symmetrizing the self-energy matrix elements with respect to the quasiparticle energies of the two states involved, i.e., by averaging $\Sigma_{pq}(\epsilon_p)$ and $\Sigma_{qp}(\epsilon_q)$. 

By contrast, the static self-energy defined in Eq.~\eqref{eq:1h+1p_static} follows a conceptually different strategy. 
Here, the matrix elements are not constructed from two separate evaluations of the self-energy. 
Instead, the frequency dependence itself is removed at the outset by evaluating each term at a single symmetrically chosen energy, typically $(\epsilon_p+\epsilon_q)/2$. 
In other words, qs$GW$ symmetrizes the matrix elements, whereas the present construction symmetrizes the energy argument entering the self-energy.

Although both approaches yield Hermitian static operators, they represent distinct approximations to the dynamical self-energy and need not coincide numerically.
In practice, however, they yield very similar results: the mean absolute and mean signed deviations between them are \SI{0.028}{\eV} and \SI{-0.012}{\eV}, respectively, with a maximum deviation of \SI{0.104}{\eV}. 
This close agreement indicates that the present static construction captures most of the effects incorporated in qs$GW$, while arising from a conceptually distinct derivation.

\begin{table*}
\caption{Core ionization energies (in eV) of the O 1s and C 1s states for the \ce{CO} molecule computed using various schemes with the aug-cc-pVTZ basis.}
\label{tab:coreIPs}
\begin{ruledtabular}
\begin{tabular}{c|cccccc}
Reference	&	\mc{3}{c}{O 1s}						&	\mc{3}{c}{C 1s}												\\
				\cline{2-4} \cline{5-7}
space		&	full dynamic	&	h\&h	&	pure static		&	full dynamic	&	h\&h	&	pure static		\\	
\hline
1h+1p		&	547.961		&	547.957	&	545.255	&	300.590	&	300.586	&	299.503	\\	
1h			&	547.958	&	547.954	&	545.245	&	300.587	&	300.582	&	299.397	\\
diagonal	&	547.957	&	547.953	&	545.234	&	300.585	&	300.581	&	299.396	\\
\end{tabular}
\end{ruledtabular}
\end{table*}

Returning to the new hierarchy introduced in this work, Table \ref{tab:coreIPs} reports the O 1s and C 1s core ionization energies of \ce{CO} obtained with the same set of schemes and the aug-cc-pVTZ basis. 
For these core ionizations, the trends observed previously become even more pronounced. 
Owing to the large energetic separation between the deep core levels and the valence orbitals, the results show only a very weak dependence on the choice of reference space. 
In particular, the diagonal approximation performs remarkably well, differing from the other dynamical schemes by only a few meV.
For the same reason, the fully dynamical and h\&h treatments yield nearly identical results. 
For core states, the effective decoupling of particle and hole branches is known to be a reliable approximation, as already exploited in early formulations of the cumulant $GW$ expansion. \cite{Lundqvist_1969,Langreth_1970,Gunnarsson_1994,Hedin_1980,Hedin_1999,Loos_2024}
In contrast, adopting a purely static scheme has a much more dramatic impact, lowering the core ionization energies by several eV. 
This reflects the crucial role of dynamical correlation effects for quantitatively accurate core-level energetics. \cite{vanSetten_2018,Golze_2018,Golze_2020,Mejia-Rodriguez_2021,Li_2022,Mukatayev_2023}

\section{Conclusion}
\label{sec:conclusion}

In this work, we have introduced a unified and systematic hierarchy of one-body Green's function methods derived from the $GW$ approximation, in which the dynamical content of the self-energy is progressively reduced systematically. 
Starting from the fully dynamical Dyson formulation, we constructed a family of approximations that interpolates between the standard $GW$ approximation and purely static effective single-particle Hamiltonians. 
Within this framework, variants that determine quasiparticle energies in the 1h+1p space, the reduced 1h or 1p spaces, or within the diagonal approximation all arise from the same supermatrix formalism through selective downfolding of the 2h1p and 2p1h configuration spaces.

This hierarchy provides a clear and physically transparent way to disentangle two effects that are usually intertwined in practical $GW$ implementations: dynamical correlation and the coupling between hole and particle sectors. 
By selectively making individual branches of the self-energy static, we have shown that frequency dependence is not a binary feature, but can instead be selectively retained or eliminated. 
In particular, the half-and-half (h\&h) schemes demonstrate that retaining dynamical effects in only one branch of the self-energy is often sufficient to reproduce the quasiparticle energies of the fully dynamical theory with high fidelity.

A key outcome of this study is the identification of numerical instabilities, rather than intrinsic physical deficiencies, as the primary origin of the previously reported failures of certain partially dynamical and non-Dyson schemes. 
The SRG-based regularization employed here removes spurious divergences in problematic energy denominators while leaving stable formulations unaffected.
Once these numerical artifacts are controlled, the h\&h approximations emerge as remarkably accurate representations of the fully dynamical $GW$ self-energy for valence IPs, with errors on the order of a few hundredths of an eV. 
In contrast, the purely static schemes remain limited by genuine physical approximations associated with the complete removal of dynamical effects, leading to systematically larger deviations, although the magnitude of these errors may be acceptable for certain applications.
Similar conclusions hold for core ionized states.
\titou{It is worth reiterating that the regularization scheme is essential to the present analysis, since without it, a reliable assessment of the different approximations would have been precluded.}

The present static construction also offers a conceptually distinct route to self-consistency. 
Although derived differently from qs$GW$, the static self-energy of Eq.~\eqref{eq:1h+1p_static} yields very similar quasiparticle energies in practice, indicating that much of the effect of qs$GW$ can be understood as a particular projection of the dynamical self-energy onto a static, Hermitian operator. 
This connection helps clarify the formal relationship between self-consistent static $GW$-like schemes and partially or fully dynamical formulations.

Beyond the specific numerical results, the main contribution of this work is conceptual. 
By embedding diagonal $G_0W_0$, Dyson supermatrix approaches, hybrid half-dynamical schemes, non-Dyson constructions, and static effective Hamiltonians within a single formal hierarchy, we provide a coherent map of the $GW$ landscape. 
This perspective not only rationalizes the performance of existing approximations but also opens the door to the design of new schemes that balance physical content, numerical robustness, and algorithmic simplicity. 
\titou{In particular, within quantum embedding frameworks, approaches that combine selective dynamical treatment with stable regularization strategies appear especially promising for extending Green's function methods to larger and more complex systems.\cite{Tolle_2021,Amblard_2022,Amblard_2023,Amblard_2024} 
In such schemes, a static low-level description can be employed for the full system, while dynamical effects are retained only in selected fragments where they are expected to play a central role. 
In this context, the present approximation could prove particularly useful in both one-body Green's function and density matrix-based embedding approaches, where reducing the dynamical content may help lower the computational cost without significantly compromising accuracy. \cite{Mejuto-Zaera_2024,Cui_2025}}

\acknowledgements{
The authors would like to thank Antoine Marie and Carlos Mejuto-Zaera for fruitful discussions.
This project has received funding from the European Research Council (ERC) under the European Union's Horizon 2020 research and innovation programme (Grant agreement No.~863481).
J.~T.~acknowledges funding from the Fonds der Chemischen Industrie (FCI) via a Liebig fellowship and support by the Cluster of Excellence ``CUI: Advanced Imaging of Matter'' of the Deutsche Forschungsgemeinschaft (DFG) (EXC 2056, funding ID 390715994).
This work used the HPC resources from CALMIP (Toulouse) under allocation 2026-18005.}

\section*{Supplementary Material}
See the \SupInf for the complete raw numerical data underlying the statistical analyses reported in the main manuscript. 

\section*{Data availability statement}
The data that supports the findings of this study are available within the article and its supplementary material.

\appendix
\section{Alternative choice for the static, Hermitian self-energy}
\label{app:ADC_choice}
Taking the particle branch as an example, one can construct a frequency-independent form of the self-energy while preserving the Hermitian character of the eigenvalue problem in different ways. 
Instead of evaluating the self-energy elements at the averaged frequency $\omega = (\epsilon_p + \epsilon_q)/2$, yielding Eq.~\eqref{eq:SigInf_2p1h}, one may alternatively symmetrize the matrix elements themselves, leading to
\begin{equation} \label{eq:ADC_sym}
\begin{split}
	\Bar{\Sigma}_{pq}^\text{2p1h} 
	& = \frac{1}{2} \qty[ \Sigma_{pq}^\text{2p1h}(\omega = \epsilon_p) + \Sigma_{qp}^\text{2p1h}(\omega = \epsilon_q) ]
	\\
	& = \frac{1}{2} \sum_{c\nu} \frac{M_{c\nu,p}M_{c\nu,q}}{\epsilon_p - \epsilon_{c} - \Om_{\nu}}
	+ \frac{1}{2} \sum_{c\nu} \frac{M_{c\nu,q}M_{c\nu,p}}{\epsilon_q - \epsilon_{c} - \Om_{\nu}}
	\\
	& = \sum_{c\nu} M_{c\nu,p}M_{c\nu,q} \frac{(\epsilon_p + \epsilon_q)/2 - \epsilon_{c} - \Om_{\nu}}{(\epsilon_p - \epsilon_{c} - \Om_{\nu})(\epsilon_q - \epsilon_{c} - \Om_{\nu})}
\end{split}
\end{equation}
This alternative construction is reminiscent of the non-Dyson ADC formulation proposed by Schirmer and co-workers. \cite{Schirmer_1998,Trofimov_2002,Schirmer_1991,Mertins_1996a,Mertins_1996b}
\titou{In the fully static limit, when the same averaging procedure is applied to both branches of the self-energy in the 1h+1p space, one recovers exactly the qs$GW$ expression [see Eq.~\eqref{eq:sigmatilde}].\cite{Faleev_2004,vanSchilfgaarde_2006,Kotani_2007}}

From a numerical perspective, we find that using the symmetrization scheme of Eq.~\eqref{eq:ADC_sym} reproduces the IPs obtained with the approach proposed in this work for the entire set considered in Sec.~\ref{sec:res}, with an accuracy better than \SI{1}{\milli\eV} for the 1h-h\&h scheme.
This numerical agreement is also valid at stretched geometries.
This confirms that the non-Dyson formulation of $GW$ is largely insensitive to the particular choice of symmetrization.

\section*{References}

\bibliography{biblio}

\end{document}